\documentclass[aps,prl,twocolumn,showpacs,preprintnumbers]{revtex4-1}
\usepackage{times}
\usepackage{graphicx,color}
\usepackage{bm}
\usepackage{amssymb,amsmath}

\begin{document}
\preprint{}
\title{Topological superconductivity in bilayer Rashba system}
\author{Sho Nakosai$^1$, Yukio Tanaka$^2$, Naoto Nagaosa$^{1,3}$}
\affiliation{
$^1$Department of Applied Physics, University of Tokyo, Tokyo 113-8656, Japan \\
$^2$Department of Applied Physics, Nagoya University, Nagoya, 464-8603, Japan \\
$^3$Cross Correlated Materials Research Group (CMRG) and Correlated Electron Research Group (CERG), ASI, RIKEN, Wako 351-0198, Japan
}

\date{\today}

\begin{abstract}
We theoretically study a possible topological superconductivity
in the interacting two layers of Rashba systems, which can be 
fabricated by the hetero-structures of semiconductors and oxides.
The hybridization, which induces the gap in the single particle dispersion, and
the electron-electron interaction between the two layers leads to the novel
phase diagram of the superconductivity. It is found that the topological 
superconductivity  {\it without breaking time-reversal 
symmetry} is realized when
(i) the Fermi energy is within the hybridization gap, and (ii) the 
interlayer interaction is repulsive, both of which can be satisfied 
in realistic systems. Edge channels are studied in a tight-binding model 
numerically, and the several predictions on experiments are also given.
\end{abstract}


\maketitle

The topological aspects of the electronic states in solids have recently
attracted intensive interest. 
In addition to the quantum Hall effect~\cite{QHE,Wen}, 
the anomalous Hall effect~\cite{AHE}, spin Hall effect~\cite{SHE}, 
and topological insulators (TI's)~\cite{TI1,TI2} turn out to be 
topological phenomena driven by the
Berry curvatures~\cite{Berry} in momentum and/or real spaces. 
Especially, the topological insulators are realized in 
the time-reversal ($T$)~symmetric systems with the 
relativistic spin-orbit interaction (SOI), and are
characterized by the $Z_2$ topological indices.
On the other hand, the concept of the topological 
superconductors (TSC) or superfluidity (TSF) has been proposed 
in the context of $^3$He~\cite{Volovik}. Recently the $p$-wave 
superconductivity with broken $T$~symmetry in Sr$_2$RuO$_4$ 
has been discussed from this respect~\cite{Maeno, Kashiwaya}.
Helical topological superconductors with $T$~symmetry have also been
discussed~\cite{Qihelicalsc}.
On the other hand, the unified scheme of the 
classification of the TIs and TSCs 
according to the three symmetries
($T$~symmetry, particle-hole symmetry, and chiral symmetry)
and the dimension of the system has been established~\cite{Schnyder}. 

The implementation or fabrication of the TSCs 
is an important issue especially because it is expected to 
support the Majorana fermions~\cite{Majorana1,Majorana2,Read,Ivanov,TSCreview}, 
a promising tool for the quantum information processes. 
An interesting recent proposal is 
the proximity-induced superconductivity of the surface state in
the three dimensional TI~\cite{FuKane}. 
Because of the electron fractionalization in the 
surface Dirac fermions, Majorana fermions are expected
to appear at the interface between the ferromagnet and 
$s$-wave superconductors both of which are on top of TI.
It is also proposed that the superconductivity in
a doped TI Cu$_x$Bi$_2$Se$_3$~\cite{Ong} can be topological~\cite{FuBerg,Sasaki,CuBiSe1,CuBiSe2,CuBiSe3}.
Besides, the possible TSC in noncentrosymmetric superconductors 
with the Rashba spin splitting has been discussed~\cite{Tanaka1,Fujimoto}. 
The basic idea 
is that the spin direction rotates with the momentum for the spin split bands, leading to the $p \pm ip$ pairing form when the superconducting order parameter is projected onto each of the bands.
However, it has been clarified that due to the interference between the two order parameters in the two spin split bands, spin triplet $p$-wave pairing should be dominant over spin singlet $s$-wave pairing for the realization of TSC~\cite{Tanaka1,NCS1,NCS2}.
One possible way to resolve this difficulty is to delete one of
the spin split bands by external magnetic field or 
by the exchange splitting due to the proximity to a ferromagnet, 
leaving alone the $p + ip$ (or $p - ip$) pairing
with broken $T$~symmetry~\cite{Fujimoto,Sau,QiQHSC,Lutchyn,Alicea,Yamakage} 
analogous to the case of Sr$_2$RuO$_4$.  
Experimentally, it is known that the superconductivity emerges
at the interface of LaAlO$_3$ and SrTiO$_3$~\cite{Reyren}, and the role of Rashba interaction has been identified there~\cite{Caviglia}. 
Therefore, the two-dimensional superconductors with Rashba coupling are now available.
It has been also reported that the superlattice of a heavy-fermion compound, CeCoIn$_{5}$, can be fabricated and its superconductivity is observed~\cite{Matsuda}.
Theoretically, a robust Fulde-Ferrell-Larkin-Ovchinnikov state has been proposed in the presence of SOI~\cite{Michaeli}. 
Also, up to now, there have been theoretical proposals about interface induced superconductivity~\cite{Yada09,Hirschfeld11} and superconductivity in multi layer system~\cite{Yanase}. 
The advantage of the interface or heterostructure systems is the great controllability of the structure, symmetry, doping, and SOI.
For example, the bilayer quantum well to form the two interacting two Rashba system can be fabricated both in the oxides~\cite{Hwang} and semiconductors~\cite{Loss}.

\begin{figure}[t]
\begin{center}
\includegraphics[width=\hsize]{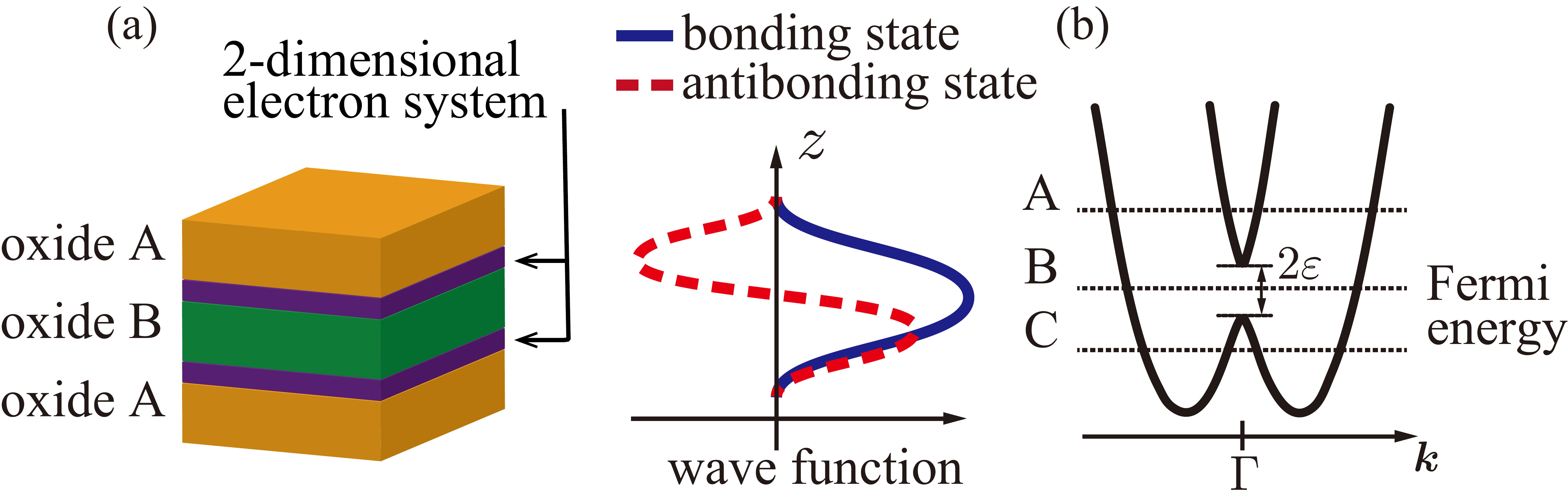}
\caption{(color online). (a) (left panel) Schematic view of the model with the interfaces between two kinds of materials A and B. An example is that A is LaAlO$_{3}$ and B is SrTiO$_{3}$. 2DEGs are formed in the SrTiO$_{3}$ side and we expect these two hybridize with sufficiently short width of the middle layer. (right panel) Schematic wave functions of bonding and antibonding states along the $z$ direction. 
The hybridization splits two Rashba bands into bonding and antibonding states with the hybridization gap $\varepsilon$. (b) Dispersion of the Hamiltonian of bilayer Rashba model Eq.~(\ref{eq:hmltn1}). The position of the Fermi energy ($A$-$C$) is crucial to nontrivial phases. The phase diagrams for each case is shown in Fig.~\ref{fig:diagram}.}
\label{fig:model}
\end{center}
\end{figure}

   In this Letter, we theoretically study the superconductivity in the interacting two layers of Rashba systems. 
In contrast to the previous proposals to delete one of the spin split bands, this system doubles the degrees of freedom without breaking $T$~symmetry, giving richer possibilities. 
The model system we consider is schematically shown in Fig.~\ref{fig:model}(a). 
At each interface, the two-dimensional electron gas (2DEG) is formed, and this sandwich structure produces two 2DEG's with the tunable interaction by the width of the middle layer. 
There are two types of the interaction. 
One is the transfer or hybridization of the electrons, and the other is the electron-electron interaction. 
The former induces a gap in the band structure, denoted by $\varepsilon$. 
The Hamiltonian for the kinetic energy is given by
\begin{equation}
\mathcal{H}_{0}=
\frac{k^2}{2m}\mathbb{I}_s\otimes\mathbb{I}_{\sigma}
-\varepsilon \mathbb{I}_{s}\otimes\sigma_{x}
+\alpha \left(k_x s_y - k_y s_x \right) \sigma_z,
\label{eq:hmltn1}
\end{equation}
where $s$ and $\sigma$ respectively denote the Pauli 
matrices for spin and layers, 
and $\alpha$ represents the strength of Rashba SOI.
The eigenvalues of this Hamiltonian is 
\begin{align}
E_{\bm{k}}=k^2/2m \pm \sqrt{\varepsilon^2+\alpha^2 k^2} =k^2/2m \pm \lambda_{\bm{k}}.
\label{eq:energy}
\end{align}

Note that these two layers are exposed to the local electric field with opposite directions, and this is described by $\sigma_{z}$ in the 3rd term of Eq.~(\ref{eq:hmltn1}).
The bands shown in Fig.~\ref{fig:model}(b) are doubly degenerate at each $k$-point in the whole Brillouin zone because this system has both $T$~symmetry, $T=is_{y}K$, where $K$ is the complex conjugation operator, and inversion ($I$)~symmetry, $I=\sigma_{x}$.
We assume the following density-density interaction between the electrons
\begin{equation}
\mathcal{H}_{\mathrm{int}}\left(\bm{x}\right)=
-U\left(n_1^2\left(\bm{x}\right)+n_2^2\left(\bm{x}\right)\right)-2Vn_1\!\left(\bm{x}\right)n_2\!\left(\bm{x}\right),
\label{eq:interaction}
\end{equation}
where $n_{\sigma=1,2}=\sum_{s=\uparrow,\;\downarrow} c^{\dagger}_{\sigma s}c_{\sigma s}$ is the electron density in layer 1 and 2.
$U$ and $V$ are intralayer interaction and interlayer one, respectively.
We construct the Bogoliubov de Gennes (BdG) Hamiltonian with mean field approximation
\begin{align}
 H_{\mathrm{BdG}}\!=\!\!
\int \! \mathrm{d}\bm{k} \Psi_{\bm{k}}^{\dagger}\!
\begin{pmatrix}
\mathcal{H}_{0}\left( \bm{k} \right) - \mu &
\Delta \\
\Delta &
-\!\left(\mathcal{H}_{0} \left( \bm{k} \right) - \mu \right)
\end{pmatrix}\!
\Psi_{\bm{k}}.
\label{eq:BdGhmltn}
\end{align}

$\tau$ is the Pauli matrices in Nambu (particle-hole) space and the basis are in the form of 
$\Psi^{\dagger}_{\bm{k}}\!=\!( c_{1 \bm{k} \uparrow}^{\dagger},\,c_{2 \bm{k} \uparrow}^{\dagger},\,c_{1 \bm{k} \downarrow}^{\dagger},\,c_{2 \bm{k} \downarrow}^{\dagger},\,c_{1 \bm{-k}\downarrow},\,c_{2 \bm{-k}\downarrow},\,-c_{1 \bm{-k}\uparrow},\,-c_{2 \bm{-k}\uparrow})$. 
Note that the BdG Hamiltonian Eq.~(\ref{eq:BdGhmltn}) 
and the analysis below is similar to that in Ref.~\cite{FuBerg}.
However, the situation we consider is quite different from theirs, i.e.,
(i) we consider the 2D case while Ref.~\cite{FuBerg} considered the 3D case,
(ii) the energy dispersion in Eq.~(\ref{eq:energy}) of the Rashba system is different from 
that of Dirac fermion due to the presence of the $k^2$ terms in the kinetic energy,
(iii) there is a tunable parameter $\alpha$ in our model, 
and (iv) we consider the layer-index while Ref.~\cite{FuBerg}  considered
the orbital index, and the meaning of $U$ and $V$ is different.
We consider pairing potential $\Delta \left(\bm{k}\right)$ according
to the lattice structure and the
interaction Hamiltonian Eq.~(\ref{eq:interaction}).
Since we are conscious of the interface between SrTiO$_{3}$ and 
LaAlO$_{3}$, the lattice symmetry is assumed to be $D_{4h}$
, which
contains following operations: a four-fold rotation about $z$ axis, a
twofold rotation about $x$ axis and a mirror reflection with respect to $xy$ plane. Note
that the mirror reflection sends a site on a layer to that on the other.
In the weak coupling limit with purely short-range
interaction, $k$ independent pairings are
favored compared with $k$ dependent anisotropic pairings.
Only 6 forms listed in Table~\ref{table:pairing} can have
nonzero values among the 16 possible products of $(1,s_x,s_y,s_z)$ and
$(1,\sigma_x,\sigma_y,\sigma_z)$.
$\Delta_1$ is the even combination of the intralayer pairings 
($c_{1\uparrow}c_{1\downarrow}\!+\!c_{2\uparrow}c_{2\downarrow} $)
and the interlayer pairing
($c_{1\uparrow}c_{2\downarrow}\!-\!c_{1\downarrow}c_{2\uparrow}$)
, and is parity even. It is
remarked that the parity operation here
represents $(\bm{r},a) \to (-\bm{r},\bar{a})$ with the layer index $a$
($\bar{a}$ represents the opposite layer to $a$), and  
is not the even-odd symmetry with respect to the exchange of the
two coordinates $(\bm{r}_1, \bm{r}_2)  \to (\bm{r}_2, \bm{r}_1)$ of
the pairing electrons. $\Delta_2$ is the interlayer 
pairing with odd parity, while $\Delta_3$ is the intralayer
with odd parity. Last, $\Delta_4$ is the
interlayer pairing with odd-parity. 

\begin{table}[t]
\begin{center}
\begin{tabular}{cccc}
\hline
\hline
\vspace{-10pt} \\
\shortstack{pairing \\ potential} & \shortstack{explicit \\ representation} & matrix & parity
\\
\hline
\vspace{-4pt} \\
$\Delta_1$: & $c_{1\uparrow}c_{1\downarrow}\!+\!c_{2\uparrow}c_{2\downarrow}, c_{1\uparrow}c_{2\downarrow}\!-\!c_{1\downarrow}c_{2\uparrow}$
& $I$, $\sigma_{x}$ & $+$
\\
$\Delta_2$: & $i(c_{1\uparrow}c_{2\downarrow}\!+\!c_{1\downarrow}c_{2\uparrow})$
& $s_{z}\sigma_{y}$ & $-$
\\
$\Delta_3$: & $c_{1\uparrow}c_{1\downarrow}\!-\!c_{2\uparrow}c_{2\downarrow}$
& $\sigma_{z}$ & $-$
\\
$\Delta_4$: & $\left( i(c_{1\uparrow}c_{2\uparrow}\!+\!c_{1\downarrow}c_{2\downarrow}),\; c_{1\uparrow}c_{2\uparrow}\!-\!c_{1\downarrow}c_{2\downarrow} \right)$
& $\left( s_{x}\sigma_{y}, s_{y}\sigma_{y} \right)$ & $-$
\\
\vspace{-10pt} \\
\hline
\hline
\end{tabular}
\caption{There are four possible nonvanishing pairing potentials in our
model with the assumption Eq.~(\ref{eq:interaction}). $\Delta_1$, $\Delta_2$, $\Delta_3$, and $\Delta_4$ belong to
 $A_{1\mathrm{g}}$, $A_{1\mathrm{u}}$, $A_{2\mathrm{u}}$, and
 $E_{\mathrm{u}}$ irreducible representations of $D_{4h}$, respectively~\cite{Ueda}. Matrix representations are off-diagonal elements of BdG Hamiltonian, i.e., entries in the column ``explicit representations" are products of $\Psi^{\dagger}$, the corresponding entries in the column ``matrix", $\tau_x$ and $\Psi$.}
\label{table:pairing}
\end{center}
\end{table}

\begin{figure}
\begin{center}
\includegraphics[width=.85\hsize]{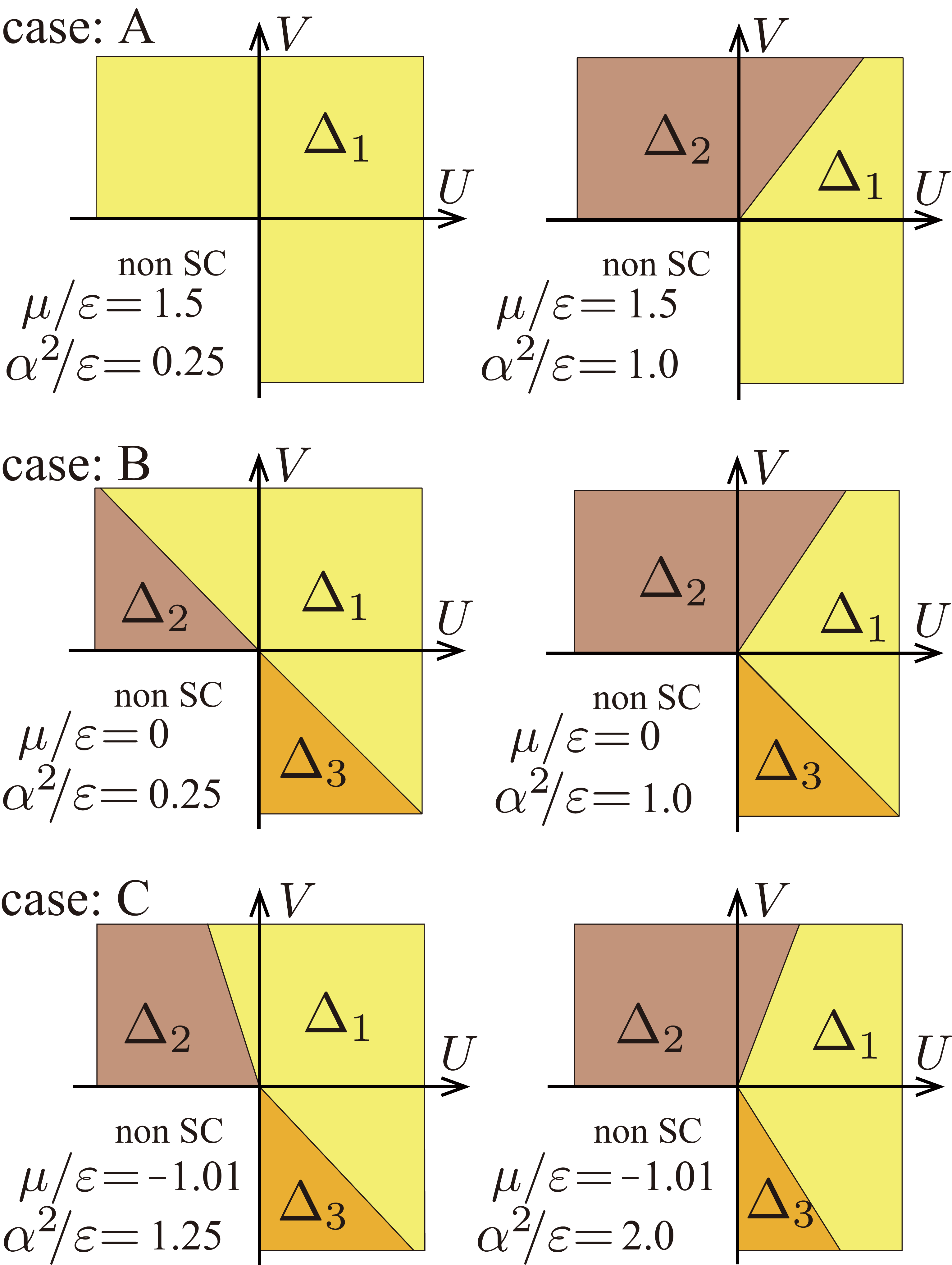}
\caption{
(color online). 
Phase diagram of superconductivity in the $UV$ plane. Yellow, brown, and
 orange indicate the region where $\Delta_{1}$, $\Delta_{2}$, and
 $\Delta_{3}$ show the strongest instability, respectively.
Here we take the unit where $m$ is the unity.
The white region is nonsuperconducting. Each row corresponds to the Fermi energy at A, B, and C in Fig.~\ref{fig:model}(b). The system can be topological only when the Fermi energy lies in the hybridization gap, i.e., case $B$, and in this case $\Delta_{2}$ region expands as the SOI increases while $\Delta_{3}$ region is independent of the strength of SOI. However, SOI is indispensable for $\Delta_{3}$. See the text.}
\label{fig:diagram}
\end{center}
\end{figure}

The excitation energy of quasiparticles are obtained by diagonalizing the
BdG Hamiltonian Eq.~(4) with fixing the pairing potential to each
$\Delta_i$ (see note~\cite{note}). We find superconducting gap for $\Delta_4$ has point
nodes (in the $k_x$ direction when one chooses $s_y\sigma_{y}$), and the others have full gap.
We estimate the superconducting critical temperature $T_{\mbox{c}}$ by analyzing superconducting susceptibility for each pairing potentials. 
The pairing susceptibility $\chi_0$ is defined as
\begin{eqnarray}
\chi_0 &=& -T\sum_{\omega_k}\sum_{\bm{k}}\mathrm{Tr}\left[
\tau_xG_0\left(\bm{k}\right)\tau_xG_0\left(\bm{k}\right)\right]
\nonumber \\
&=&-\sum_{\bm{k}}\!\left[\frac{1-2f\left(\xi_{\bm{k}}+\lambda_{\bm{k}}\right)}
{2\left(\xi_{\bm{k}}+\lambda_{\bm{k}}\right)}
+\frac{1-2f\left(\xi_{\bm{k}}-\lambda_{\bm{k}}\right)}
{2\left(\xi_{\bm{k}}-\lambda_{\bm{k}}\right)}\right],
\label{eq:chi0}
\end{eqnarray}
where $f(E)$ is the Fermi distribution function and $\xi_{\bm{k}}=k^2/2m-\mu$.
Hereafter, we assume the weak coupling limit.
The other susceptibilities can be calculated by replacing $\tau_{x}$ with $\tau_{x} s \sigma$ ($s \sigma$ is a matrix representation for each pairing in Table~\ref{table:pairing}).
As a result, they can be expressed by $\chi_0$, which contains the logarithmic divergence estimated by the value at the Fermi surface.
The linearized gap equations are obtained as (two of them are shown explicitly)
\begin{eqnarray}
\begin{vmatrix}
U\chi_{0}-1\!\! & \!\!U\chi_{01} \!\\
V\chi_{10}\!\! &\!\! V\chi_{1}-1\!
\end{vmatrix}
\!\!=\!\!
\left[U+V\left(\varepsilon/\lambda_{\bm{k}}\right)^2\right]\chi_{0}\!\!-\!\! 1 \!\!=\!\! 0
\hspace{6pt}&\mathrm{for}&\;\Delta_{1},\hspace{20pt}
\\
V\chi_{2} \!=\!
V\left[1-\left(\varepsilon/\lambda_{\bm{k}}\right)^2\right]\chi_{0}
\!=\!1
\hspace{6pt}&\mathrm{for}&\;\Delta_{2},\hspace{20pt}
\end{eqnarray}
when the Fermi surface lies in the hybridization gap (case $B$).
Note that, for $\Delta_{1}$, there are two choices of vertex, $\tau_{x}I$ or $\tau_{x}\sigma_{x}$. 
Correspondingly, the gap equation has $2\times2$ form with the coefficients 
$\chi_0\!=\!-\int\mathrm{d}\xi D_{+}(\eta)(\tanh(\beta\eta/2)/2\eta)$,
$\chi_{01}\!=\!\chi_{10}\!=\!(\varepsilon/\lambda_{\bm{k}})\chi_{0}$ and
$\chi_{1}\!=\!(\varepsilon/\lambda_{\bm{k}})^2\chi_{0}$.
In the linearized gap equations for $\Delta_3$ and $\Delta_4$, the
susceptibilities are
$\chi_3=2\chi_4=[1-(\varepsilon/\lambda_{\bm{k}})^2]\chi_0$ and
the coefficients are $U$ and $V$, respectively.
$\lambda_{\bm{k}}$ is evaluated at the momentum crossing the Fermi energy since we assume the weak coupling limit.
$D_{+}(\eta)$ is the density of states of the outer cone in the band structure [Fig.~\ref{fig:model}(b)] and $\eta$ is energy measured from the Fermi energy.
From the highest $T_c$, we obtain the phase diagram in the $UV$ plane as
shown in Fig.~\ref{fig:diagram} (case $B$). 
It is seen that the phase diagram consists of $\Delta_1$, $\Delta_2$, $\Delta_3$, and nonsuperconducting regions. 
$\Delta_4$ cannot be the leading instability in this phase diagram, and
hence all the pairing states are fully gapped.
The phase boundaries depend only on $U/V$ since they are determined by the comparison
among the effective interactions for each pairing, which contain terms
linear in $U$ and/or $V$.
When both $U$ and $V$ are repulsive, the superconducting instability is absent, while some pairing occurs in all the other cases. 
When the intralayer attraction $U(>0)$ is dominant, the conventional pairing $\Delta_1$ occurs, while the unconventional $\Delta_2$ and $\Delta_3$ are realized in other regions. 
Starting from the single layer case where the superconductivity is realized as in the case of LaAlO$_3$/SrTiO$_3$ interface~\cite{Reyren}, 
$U$ is considered to be attractive. Assuming that this attractive interaction $U$ is
given by the short-range electron-phonon coupling reduced by the Coulomb repulsive interaction, the repulsive interlayer $|V|$ is expected to be larger than $U$. 
Therefore, we expect the situation where $\Delta_3$ is realized in the bilayer Rashba system. 
Note that although the region of $\Delta_3$ is independent of the Rashba SOI, the effective coupling constant for $\Delta_3$ is $U[1-(\varepsilon/\lambda_{\bm{k}})^2]$ which is zero for $\alpha=0$. 
Therefore, $\Delta_3$ is the superconductivity induced by Rashba SOI~\cite{Trauzettel}.
When there is no SOI, the bands are simply bonding and antibonding ones with the hybridization gap, where the interactions $U$ and $V$ drive only conventional spin singlet pairing. 
The spin states are mixed with finite SOI, and this causes unconventional spin triplet pairing. 
According to the symmetry of pairing, $U$ drives $\Delta_1$ and $\Delta_3$ while $V$ drives $\Delta_1$, $\Delta_2$ and $\Delta_4$. 
These features are valid for both cases with single and multi Fermi surfaces. 
Additional Fermi surface in case $A$ and $C$ gives another logarithmically divergent term and modifies phase boundaries. 
In case $A$, $\Delta_3$ instability is always weaker than that of $\Delta_1$, and $\Delta_2$ instability is also weaker when SOI is less than a certain critical value. 
In case $C$, the boundary between $\Delta_1$ and $\Delta_3$ is determined by the strength of SOI. Note that case $C$ can appear only when SOI is strong enough for the band structure to have double minima [see Fig.~\ref{fig:model}(b)]. 
There can be rapid change of phase boundaries between case $A$($C$) and $B$ because the density of states in 2D has step-function-like behavior at the edge of band bottom or top, which gives an additional logarithmic singularity. 
Apart from topological nature, the fact that various kinds of pairing potential occur corresponding to the strength of SOI is an interesting result in this Letter.

Next we consider the topological nature of each pairing state. 
According to the generic classification scheme~\cite{Schnyder},
the present system belongs to DIII in 2D. Therefore, the homotopy class
is characterized by $Z_2$ number in contrast to the $Z$ in 3D.
We can classify the superconductors into two categories, i.e.,
topologically trivial superconductors and the helical one. 
Since our system has $T$~symmetry, particle-hole symmetry, 
and $I$~symmetry in the normal state Hamiltonian Eq.~(\ref{eq:hmltn1}), the criterion in Refs.~\cite{FuBerg, Sato} can be applied.
Namely, the present system becomes a topological superconductor if (i) it 
has odd-parity pairing symmetry with full superconducting gap ($\Delta_{2}$ or $\Delta_{3}$) 
and (ii) odd number of Kramers pairs at the time-reversal invariant momenta 
$\Gamma_{\alpha}$'s are below the Fermi surface. 
The condition is explicitly written in
\begin{align}
\left( -1 \right)^{\nu}
=\prod_{\alpha} \left( -1 \right)^{N\left(\Gamma_{\alpha}\right)},
\label{eq:z2}
\end{align}
where $N(\Gamma_{\alpha})$ is the number of occupied Kramers pairs at $\Gamma_{\alpha}$.
To satisfy these conditions in our system, the Fermi surface must lie in the hybridization gap, i.e., case $B$. 
In terms of the symmetry of the pairing potential, there are two possibilities, $\Delta_{2}$ or $\Delta_{3}$, for TSC.

Next, we confirm the system is exactly in the topological phase under such 
conditions. It is known that a nontrivial topological number in a bulk state 
accompanies topologically protected states localized at an edge in a finite system 
-- bulk-edge correspondence. We study following tight-binding model 
describing continuous model Eq.~(\ref{eq:BdGhmltn}) in the low energy regime
\begin{align}
H\!=\!\sum_{\bm{r}\bm{r}^{\prime}}c^{\dagger}_{\bm{r}}t_{\bm{r}\bm{r}^{\prime}}c_{\bm{r}^{\prime}}
\!+\!
\left( \!\sum_{\bm{r}}\!c^{\dagger}_{\bm{r}}\Delta c^{\dagger}_{\bm{r}} \!-\!
\mathrm{h. c. } \right)
\!-\!\sum_{\bm{r}}(\mu^{\prime}+\varepsilon \sigma_x)c^{\dagger}_{\bm{r}}c_{\bm{r}},
\label{eq:tightbinding}
\end{align}
\begin{align}
t_{\bm{r}\bm{r}^{\prime}} =
\begin{cases}
-t\pm it^{\prime}s_{y}\sigma_{z}
& \mathrm{for\ \ }
\bm{r}\!=\!\bm{r}^{\prime}\!\pm\! a_{x}\bm{e}_{x}
\\
-t\mp it^{\prime}s_{x}\sigma_{z}
& \mathrm{for\ \ }
\bm{r}\!=\!\bm{r}^{\prime}\!\pm\! a_{y}\bm{e}_{y}
\end{cases}
.
\label{eq:hopping}
\end{align}

\begin{figure}
\begin{center}
\includegraphics[width=\hsize]{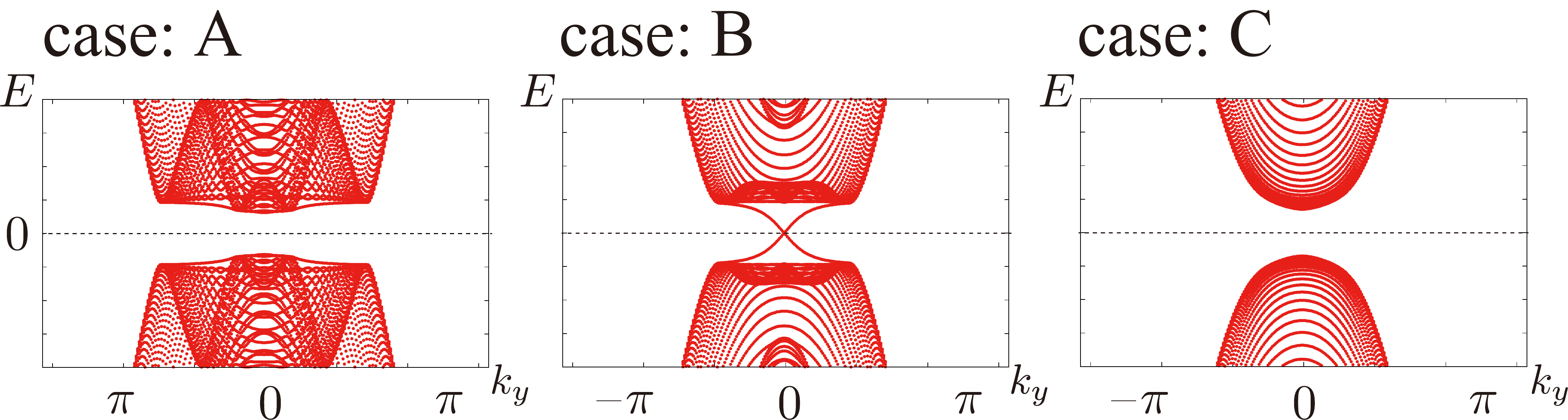}
\caption{(color online). 
Energy spectrum of the cylindrical sample with $\Delta_3$ pairing for different positions of the 
Fermi surface, i.e., case $A$, $B$ and $C$ (from left to right); see Fig.~\ref{fig:model}(b). 
These figures show that helical edge modes appear at each edge of the sample in the superconducting gap only when the Fermi surface lies within the hybridization gap.}
\label{fig:spectrum}
\end{center}
\end{figure}

The hopping parameters are related to the ordinary parameters in 
Rashba system in the form of $t=-1/2m$, $t^{\prime}=\alpha/2$ and 
$\mu^{\prime}=\mu-2/m$.
We consider the $\Delta_3$ pairing state in the cylindrical sample, i.e., open boundary condition to the $x$ direction and periodic one to the $y$ direction.
The energy spectrum of this model is shown in Fig.~\ref{fig:spectrum} with 
different positions of the Fermi surface. One can see that there are states 
crossing bulk superconducting gap only when the Fermi surface lies 
within the hybridization gap (case $B$). The right-going and left-going states are 
doubly degenerate, respectively, and localized at opposite edges.
These are nothing but the helical Majorana edge modes.  
Then, we can conclude that when the bulk has nontrivial $Z_2$
number, corresponding helical edge states appear at the edge of the system. 
Helical edge states are also generated in the case of $\Delta_2$.
However, there are no edge states with $\Delta_1$. These results are totally consistent with the topological nature of the bulk states.

This helical superconductivity has several unique features compared 
with the chiral one~\cite{Tanaka1, Shindou, Asano, Beri}.
Since the $T$~symmetry is preserved, 
there remains the Kramer's degeneracy for the states. Therefore,
the spin degrees of freedom is still alive for the Majorana edge channels, 
and Majorana bound state in the core of the vortex.
This fact leads to the several nontrivial physical consequences. 
One possible experiment is the Andreev reflection at the
interface of the normal metal and the helical superconductor~\cite{Tanaka1}.
Because of the two counter propagating Majorana edge channels with 
opposite spins, the spin conductance is strongly dependent on the 
incident angle of the electrons, and also very sensitive to the 
external magnetic field $B$. Since $B$ acts mostly through the 
Doppler shift,  even a tiny $B$ can induce the large spin conductance
of the order of the charge conductance~\cite{Tanaka1}.
Magnetic impurity at the helical edge channels is another
interesting possibility~\cite{Shindou}. Because of the
Majorana nature, the spin components are reduced
from 3 to 1, i.e., it is of Ising-like. This fact leads to
the novel Kondo effect of the magnetic impurity
at the edge of a helical superconductor showing the 
anisotropic and strong temperature dependent 
magnetic susceptibility. 
Josephson junction between the two helical superconductors 
offers a laboratory for the even more rich physics~\cite{Asano}.
Because the forward scattering channels are open with the
4 species of Majorana channels, Tomonaga-Luttinger 
liquid can be realized in the two interacting sets of the 
helical edge channels at the Josephson junction. 
This leads to the novel temperature and bias dependence of 
the quasiparticle tunneling conductance through this junction~\cite{Asano}.
Last, the appearance of the two zero energy 
Majorana bound states at the core of the vortex is expected~\cite{Teo,Roy},
but the detailed analysis of this subject is left for future studies.

In summary, we have studied the possible topological superconductivity in the
interacting bilayer Rashba systems, which can be realized at the interfaces 
of oxides and semiconductors. When the Fermi energy is in the hybridization gap 
and the interlayer interaction is repulsive
and stronger than the attractive intralayer interaction,
the topological superconductivity is realized. 
The helical Majorana edge channels are confirmed numerically 
in a tight-binding model in this case.
Andreev reflection, Josephson junctions, and the Kondo effect are 
the possible phenomena to demonstrate the 
unique features of the helical Majorana modes which 
have spin degrees of freedom.

The authors acknowledge the fruitful discussion with Harold~Y.~Hwang and Bjorn Trauzettel. 
This work is supported by Grant-in-Aid for Scientific Research
(Grants No. 17071007, No. 17071005, No. 19048008,
No. 19048015, No. 22103005,
No. 22340096, and No. 21244053) 
from the Ministry of Education, Culture,
Sports, Science and Technology of Japan, Strategic
International Cooperative Program (Joint Research Type)
from Japan Science and Technology Agency, and Funding
Program for World-Leading Innovative RD on Science and
Technology (FIRST Program).
%
%
%

%
%

\begin{thebibliography}{51}
\bibitem [{not()}]{note}%
  \BibitemOpen
  \href@noop {} {}\bibinfo {note} {The energy dispersion for the Hamiltonian
  with fixing the pairing potential to each $\Delta_i$: \\
  $E_1\!\!=\!\pm\!\sqrt{\xi^2\!\!+\!\!\alpha^2
  k^2\!\!+\!\!\varepsilon^2\!\!+\!\!\Delta_1^2\!\pm\!2\xi\sqrt{\alpha^2
  k^2\!\!+\!\!\varepsilon^2}\,}$ ,\\
  $E_2\!\!=\!\pm\!\sqrt{\xi^2\!\!+\!\!\alpha^2
  k^2\!\!+\!\!\varepsilon^2\!\!+\!\!\Delta_2^2\!\pm\!2\sqrt{(\alpha^2
  k^2\!\!+\!\!\varepsilon^2)\xi^2\!\!+\!\!\varepsilon^2\Delta_2^2}}$, \\
  $E_3\!\!=\!\pm\!\sqrt{\xi^2\!\!+\!\!\alpha^2
  k^2\!\!+\!\!\varepsilon^2\!\!+\!\!\Delta_3^2\!\pm\!2\sqrt{(\alpha^2
  k^2\!\!+\!\!\varepsilon^2)\xi^2\!\!+\!\!\varepsilon^2\Delta_3^2}}$ , and \\
  $E_4\!\!=\!\pm\!\sqrt{\xi^2\!\!+\!\!\alpha^2
  k^2\!\!+\!\!\varepsilon^2\!\!+\!\!\Delta_4^2\!\pm\!2\sqrt{(\alpha^2
  k^2\!\!+\!\!\varepsilon^2)\xi^2\!\!+\!\!(\alpha^2
  k_x^2\!\!+\!\!\varepsilon^2)\Delta_4^2}}$ , where $\xi=\xi_{\bm{k}}$ is
  defined just below Eq.(\ref{eq:chi0})}\BibitemShut {NoStop}%
\end{thebibliography}

\end{document}